# Enhanced Light Emission from the Ridge of Two-dimensional InSe Flakes

Yang Li[1§], Tianmeng Wang[1§], Han Wang[2§], Zhipeng Li[1,3], Yanwen Chen[1], Damien West[2], Raman Sankar[4,5], Rajesh K. Ulaganathan[4,5], Fangcheng Chou[4,5], Christian Wetzel[2], Cheng-Yan Xu[6*], Shengbai Zhang[2*], Su-Fei Shi[1,7*]

[1.] Department of Chemical and Biological Engineering, Rensselaer Polytechnic Institute, Troy, NY 12180
[2.] Department of Physics and Astronomy, Rensselaer Polytechnic Institute, Troy, NY 12180
[3.] School of Chemistry and Chemical Engineering, Shanghai Jiao Tong University, Shanghai, 200240, China
[4.] Institute of Physics, Academia Sinica, Nankang, Taipei, Taiwan 11529
[5.] Center for Condensed Matter Science, National Taiwan University, Taipei, Taiwan 10617
[6.] School of Materials Science and Engineering, Harbin Institute of Technology, Harbin 150001, China
[7.] Department of Electrical, Computer, and Systems Engineering, Rensselaer Polytechnic Institute, Troy, NY 12180

[§] These authors contributed equally to this work
[*] Corresponding authors: shis2@rpi.edu, zhangs9@rpi.edu, cy_xu@hit.edu.cn

**Abstract**

InSe, a newly rediscovered two-dimensional (2D) semiconductor, possesses superior electrical and optical properties as a direct bandgap semiconductor with high mobility from bulk to atomically thin layers, drastically different from transition metal dichalcogenides (TMDCs) in which the direct bandgap only exists at the single layer limit. However, absorption in InSe is mostly dominated by an out-of-plane dipole contribution which results in the limited absorption of normally incident light which can only excite the in-plane dipole at resonance. To address this challenge, we have explored a unique geometric ridge state of the 2D flake without compromising the sample quality. We observed the enhanced absorption at the ridge over a broad range of excitation frequencies from photocurrent and photoluminescence (PL) measurements. In addition, we have discovered new PL peaks at low temperature due to defect states on the ridge, which can be as much as ~ 60 times stronger than the intrinsic PL peak of InSe. Interestingly, the PL of the defects is highly tunable through an external electrical field, which can be attributed to the Stark effect of the localized defects. InSe ridges thus provide new avenues for manipulating light-matter interaction and defect-engineering which are vitally crucial for novel optoelectronic devices based on 2D semiconductors.

**Main text**

Two-dimensional (2D) semiconductors have attracted intense research interest because of their well-defined bandgap and atomic scale thickness, promising optoelectronics applications beyond the scaling limit.[1-3] As the most studied 2D semiconductors, transition metal dichalcogenides (TMDCs) undergo an indirect to direct bandgap transition when they are thinned down to the monolayer limit.[4,5] Furthermore, monolayer TMDCs (such as $MoS_2$, $WSe_2$, etc) exhibit rich exciton physics and possess additional valley degree of freedom.[6-8] However, the relatively low carrier mobility and low absorption associated with a monolayer impede the development of electronic and optoelectronic application based on TMDCs.[9-12]

In contrast, bulk InSe has a direct bandgap at the $Z$ point of the Brillouin zone,[13-16] and its mobility can be as high as ~ 20,000 $cm^2/Vs$.[14] Interestingly, InSe retains both the direct bandgap feature and high carrier mobility down to a few atomic layers, rendering it a promising candidate for nanoscale high-speed optoelectronic applications.[17-20] However, the orbital character of the band edges are distinct from TMDs, leading to altered optical selection rules. In particular, the CBM of InSe is dominated by the $s$ orbital of the In atom and the first VBM is dominated by the $p_z$ oribtal of the Se atom (Fig. 1b).[15,21] As a result, the absorption dipole is out-of-plane and cannot be directly excited by a normally incident light at resonance, prohibited by the selection rule ($E \perp c$ is forbidden as shown in Fig. 1b).[22-25] This forbidden transition for normally incident excitation, only allowed by mixing with the deeper valence bands with Se $p_{xy}$ character or by excitation with excess photon energy, hinders potential nanoscale optoelectronic applications based on InSe in which a response to normally incident light is often required. Previously, InSe flakes were coupled to a silica-nanoparticles-roughened substrate to increase the coupling to the normally incident light, resulting in the enhancement of both light absorption and emission.[24]

In this work, we exploit a unique topology feature, naturally-formed ridges in a 2D InSe flake, to address this challenge by coupling the electrical field of the incident light with the out-of-plane dipole moment. The InSe ridge in a 2D flake, schematically shown in Fig. 1a and b, allows the optical field of the normally incident light to couple with the originally in-plane dipole. We employed photoluminescence (PL) spectroscopy and scanning photocurrent microscopy (SPCM) to reveal the emission/absorption enhancement. In addition, we observed new PL peaks at low temperature due to defect states at the ridge, which can be ~ 60 times stronger than the intrinsic PL of InSe. Interestingly, the PL of the defects is highly tunable by a gate voltage through the Stark effect, which is attributed to the localized nature of the defects.

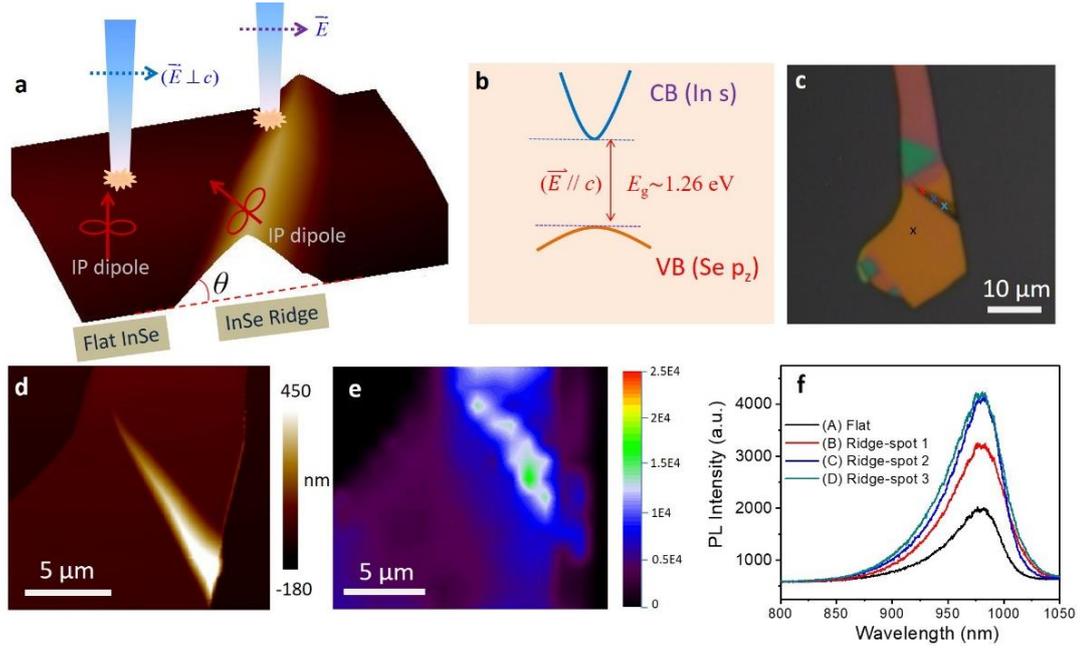

**Figure 1. PL enhancement from the ridge of a 2D InSe flake.** (a) A schematic of the in-plane (IP) emission dipole orientation with the normally incident light exciting both the flat and ridge region of the InSe flake. $\theta$ denotes the angle of the ridge. (b) A schematic of the bandstructure of bulk InSe at the $Z$ point. The first interband transition is dominantly induced by the optical field parallel to the c axis of InSe. (c) Optical microscope image and (d) AFM topography of an InSe flake with a ridge. (e) The spatially-resolved PL mapping of the InSe flake containing the ridge shown in (d). (f) The PL spectra from the positions in (c) show enhanced PL at the ridge compared to that from the flat region.

InSe was mechanically exfoliated on a Si/SiO$_2$ substrate with an oxide layer of 300 nm, and large flakes with ridges (Fig. 1c) were often found. From the atomic force microscopy (AFM) topography, we determined the thickness of the flake in Fig. 1d to be approximately 80 nm, and the height and width of the ridge are about 350 nm and 2 μm, respectively. This corresponds to a $\theta$ (Fig. 1a) of 19.3 degree. We have measured more than 20 ridges, and found that the average angle ($\theta$) of the ridge varies from 10 to 30 degree (See Fig. S1 in Supporting Information). We further carried out the spatially-resolved PL mapping of the flake at room temperature. As shown in Fig. 1e, the PL from the flat region is relatively homogenous, while that from the ridge region shows inhomogeneity and is evidently enhanced. Detailed PL spectra from different spots on the ridge is shown in Fig. 1f, in which PL from the ridge is enhanced by ~ 2.6 times compared to that from the flat region (black curve in Fig. 1f). We note, however, that the PL spectra from both the ridge region and flat region are centered at ~ 985 nm (~ 1.26 eV). This corresponds to the transition from the first VBM to CBM at $Z$ point as shown in Fig. 1b, and is consistent with previous reports of 2D InSe.[15,17,24] Similar PL enhancement has been observed in more than twenty InSe ridges that we have investigated, with the thickness varying from ~25 to ~120 nm (Part II in SI).

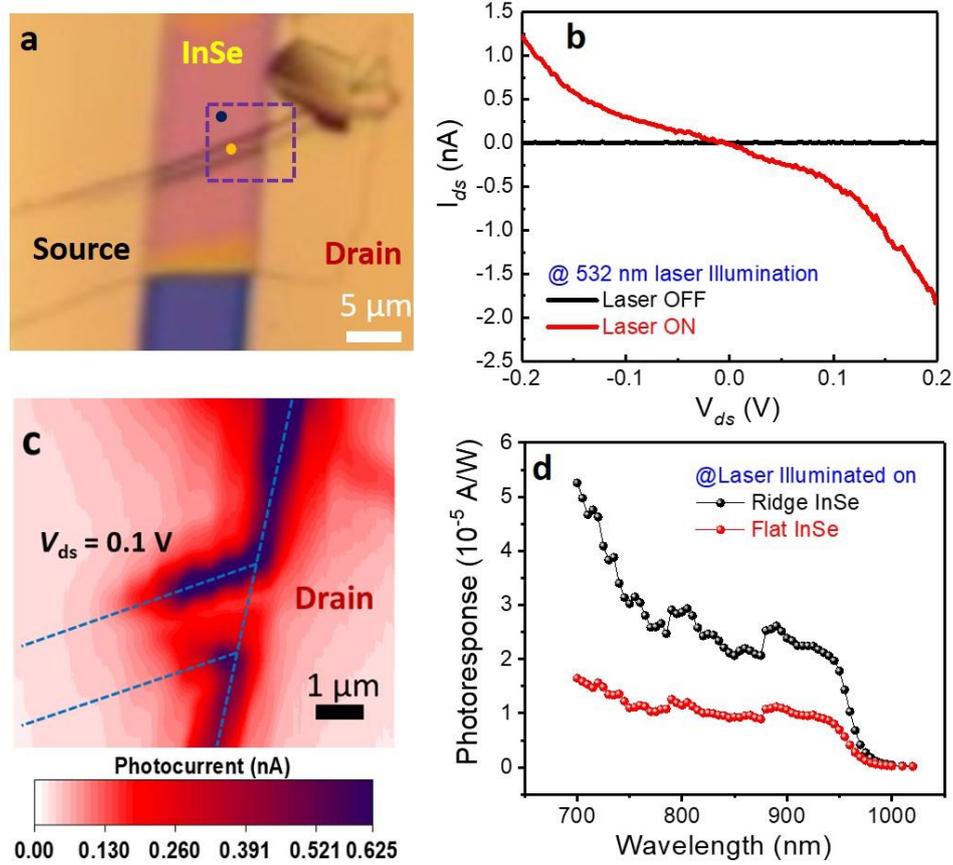

**Figure 2. Photocurrent enhancement from the InSe ridge.** (a) Optical image of a multi-layer InSe device for photocurrent measurements. The thickness of the flake is about 60 nm. (b) *I-V* curves of the InSe device with or without CW laser (λ=532 nm) illuminating on the ridge. The yellow (black) circle in Fig. 2a is the laser illumination spot on the InSe ridge (flat InSe), and the excitation power is 10 μW. (c) Spatially resolved photocurrent image of the device with the source-drain voltage of 100 mV and the excitation power of 10 μW. The blue dash lines outline the edges of the ridge, and the yellow dash line is the contact interface between Au and InSe. The scanning area is shown as the square in (a). (d) Photocurrent at the bias voltage of 100 mV as a function of excitation wavelength on the flat and ridge region of the InSe flake.

We performed photocurrent measurements of the devices based on the InSe ridge, with the consideration that the photocurrent is proportional to the absorption of the semiconductor.[26-28] The device was fabricated by evaporating gold electrodes using a shadow mask to a pre-identified few-layer InSe flake with a ridge, avoiding possible deformation or contamination of the ridge induced by photolithography or e-beam lithography. A typically device is shown in Fig. 2a, and the few-layer InSe flake is about 60 nm thick. The electrical transport property was characterized with I-V measurements at 77 K (Fig. 2b) with and without light illumination on the ridge. The device shows a low dark current but large photocurrent response to the light, with on/off ratio exceeding 1,000 at the bias voltages of 200 mV and –200 mV. This sensitive response to light can be exploited for applications in photodetectors.

The spatially-resolved photocurrent was obtained by scanning photocurrent microscopy (SPCM) measurements to distinguish the different response from the flat and ridge region of the InSe flake at 77 K, with a CW laser excitation centered at 532 nm. Fig. 2c shows the spatially resolved photocurrent response at a bias voltage of 100 mV near the right electrical contact. The photocurrent near the left electrical contact at 100 mV and photocurrent mapping at –100 mV are shown in Supporting Information (Part III). By comparing the SPCM maps with the reflection images obtained simultaneously (See Fig. S7 and S8 in SI), it is evident that the photocurrent response is dominated by the contribution from the contact and along the ridge area. The enhanced photocurrent from the Au/InSe interface can be attributed to the enhanced driving force, originating from the built-in field at the junction. The enhancement from the ridge, however, is of a different nature. With the absence of the built-in field, the enhancement of photocurrent is due to the enhancement of photoconductivity, which originates from the absorption enhancement. This interpretation is confirmed in Fig. S7 and S8 in SI, in which the photocurrent from the ridge (and the flat region) changes sign as the bias voltage switches from positive to negative. To further investigate the nature of the enhanced photo-conductivity at the ridge, we measured the photoconductivity enhancement as a function of the excitation light wavelength. As shown in Fig. 2d, the photocurrent from both the flat and the ridge region of the InSe flake exhibit similar band edge rising behavior as the exciting wavelength is shorter than ~ 975 nm, consistent with our observation of the PL spectra in Fig. 1f. Once above the band edge, the photocurrent on the ridge is enhanced over a broad range of excitation wavelength (700 to ~950 nm) by 2-3 times.

Both the PL enhancement and photoconductivity enhancement of the InSe ridge can be explained with the absorption enhancement. Per our previous discussion of the out-of-plane dipole, the enhancement of absorption ($\eta$) on the ridge originates from the increased out-of-plane component of the optical field, and it can be expressed as

$$\eta = \alpha_{E_\perp} \cos\theta + \alpha_{E_\parallel} \sin\theta, \quad (1)$$

where $\eta$, $\alpha_{E_\perp}$, $\alpha_{E_\parallel}$, and $\theta$ are effective absorption on the ridge, absorption coefficient for E⊥c, absorption coefficient for E//c, and incident angle determined by AFM measurement of the ridge. Previous first-principle calculations have reported the ratio of $\alpha_{E_\parallel}/\alpha_{E_\perp}$ at excitation wavelength of 532 nm to be ~ 6.[24] Considering that the incident angle $\theta$ in Figure 2a is ~ 22 degrees, the predicted absorption anisotropy of 6 will lead to enhanced power absorption of 3.2, which is in good agreement with our observation of the enhanced PL and photoconductivity. Additional PL enhancement from improved coupling of the emission dipole to free space was not observed. This lack of additional enhancement is partly due to the excessive energy of the excitation photon which allows for phonon-assisted recombination, relaxating the restriction of the selection rule and partly due to the large NA of the objective (~0.6) which collects emitted photons with a large solid angle.

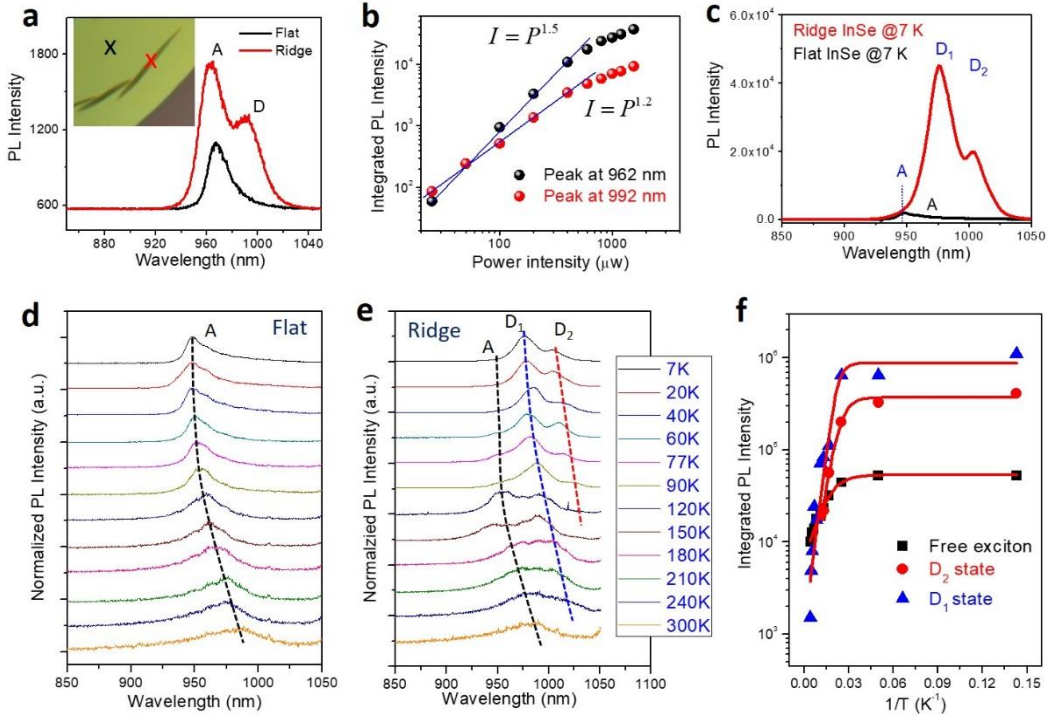

**Figure 3. Enhanced PL from defects on the InSe ridge.** (a) PL spectra of the flat (dark X) and ridge (red X) region of a InSe flake at 77 K. Inset: optical image of the InSe flake with ridges. (b) Integrated PL intensity as a function of the excitation power for the two peaks at 960 and 995 nm at 77 K. (c) PL spectra of the flat InSe and InSe ridge at 7 K. (d) and (e) Temperature dependent PL spectra of the flat InSe and InSe ridge from 7 to 300 K, respectively. A represents the photon emission from the free exciton, while $D_1$ and $D_2$ are from emission due to the newly formed defects. PL spectra in (c) and (d) are renormalized for comparison. (f) The integrated PL intensity of A, $D_1$ and $D_2$ peaks as a function of reciprocal temperature from 7 to 300 K. The integrated PL intensity of the A exciton dependence on the temperature is extracted from the flat InSe, where those of $D_1$, $D_2$ are extracted from ridge InSe.

Interestingly, new PL peaks emerge on InSe ridge at low temperature. As shown in Fig. 3a, a new PL peak (labelled as D) centered at ~ 995 nm emerges on the InSe ridge at 77 K, along with the intrinsic exciton PL peak (A) centered at ~ 960 nm which is the same as the PL from the flat InSe (black curve in Fig. 3a). The excitation power dependence of the new PL peak (D) is different from the intrinsic PL of flat InSe (A). We plot the integrated PL signal from the ridge as a function of the excitation power in Fig. 3b (also S10 in SI). PL intensity from the A and D peaks depend on power nonlinearly as ~ $P^\alpha$, where $P$ is the excitation power. The fitting of the experimental data shows that $\alpha$ is ~ 1.5 for the A peak and ~ 1.2 for the D peak. The obtained superlinear power dependence of the A peak with the exponent of ~ 1.5 suggests that the PL of peak A at 77 K has a dual nature of free carrier and excitonic behavior, due to the small exciton binding energy. The power exponent of peak D clearly differs from that of the peak A and is closer to 1.

Cooling the InSe sample down to lower temperature introduces even more striking changes to the PL spectra from the InSe ridge. Fig. 3c shows the PL spectra of flat InSe and InSe ridge at 7 K. It is observed that not only does the new PL peak (D) split into two peaks ($D_1$ and $D_2$), but the PL intensities of the $D_1$ and $D_2$ peaks are drastically enhanced compare to that of A peak from the ridge, with the enhancement factor of ~ 60 for $D_1$ and ~ 20 for $D_2$. The PL intensities from the defects are also greatly enhanced compared to that from the flat region (black curve in Fig. 3c). Detailed temperature evolution of the PL from flat InSe and the InSe ridge are shown as normalized PL in Fig. 3d and 3e. PL from the flat InSe flake (Fig. 3d) shows a blue shift of the exciton peak A as the temperature is decreased, which can be described by the empirical Varshni relation as in traditional semiconductors.[29]

The integrated intensities of A, $D_1$ and $D_2$ peaks as a function of temperature are shown in Fig. 3f. The nearly two orders of magnitude increase of PL intensity of the $D_1$ peak vs the A peak is evident at 7 K. The intensity of all three peaks are associated with an activation behavior as the temperature changes and can be fitted by an Arrhenius equation (Fig. 3f): $I(T) = \dfrac{I_0}{1 + A\exp(-E_B/k_B T)}$, where $I_0$, $E_B$ and $k_B$ are the PL intensity at 0 K, the thermal activation energy, and Boltzmann constant, respectively. The binding energy of A peak is estimated to be 16.4 meV, which is close to the previously reported value of exciton binding energy in bulk InSe (14.5 meV).[30-33] Arrhenius fitting of the peak $D_1$ and $D_2$ results in the defect binding energy of 35.3 meV ($D_1$) and 25.6 meV ($D_2$), respectively. The larger binding energy from the defects results in the red shift of the $D_1$ and $D_2$ peak compared with that of the A peak. The small binding energy of the A peak indicates that the band edge transition dominates the PL at room temperature.

The activation behavior of the PL suggests that the new peaks are from defects. But why would defects form at the ridge? A close examination of the AFM topology of the InSe ridge (Fig. 1d) suggests that the geometry will induce a strain as high as ~8.8% in the InSe flake if there is no mechanical break-down.[34,35] This large strain could mechanically break bonds locally at the ridge, which releases the strain and introduces defect states. This is supported by our separate study of strain effect on InSe flake, which revealed an extremely sensitive strain-induced bandgap change rate (~ 150 meV/% strain), without the appearance of additional PL peaks.[36] This hypothesis is also supported by our control experiments in which we intentionally introduce strain into InSe by forming wrinkles in an InSe flake, which is fabricated through exfoliation on a strained PDMS film, followed by releasing the strain in PDMS (see SI). PL from the InSe winkles show a significant peak shift even at room temperature. PL from the InSe ridge at low temperature, however, exhibits the intrinsic A peak at the same wavelength (emission photon energy) as the PL from the flat InSe, indicating the absence of strain in the ridge. The additional peaks thus can be attributed to the newly formed defects, with the activation energies given by the previous fitting with the Arrhenius equation.

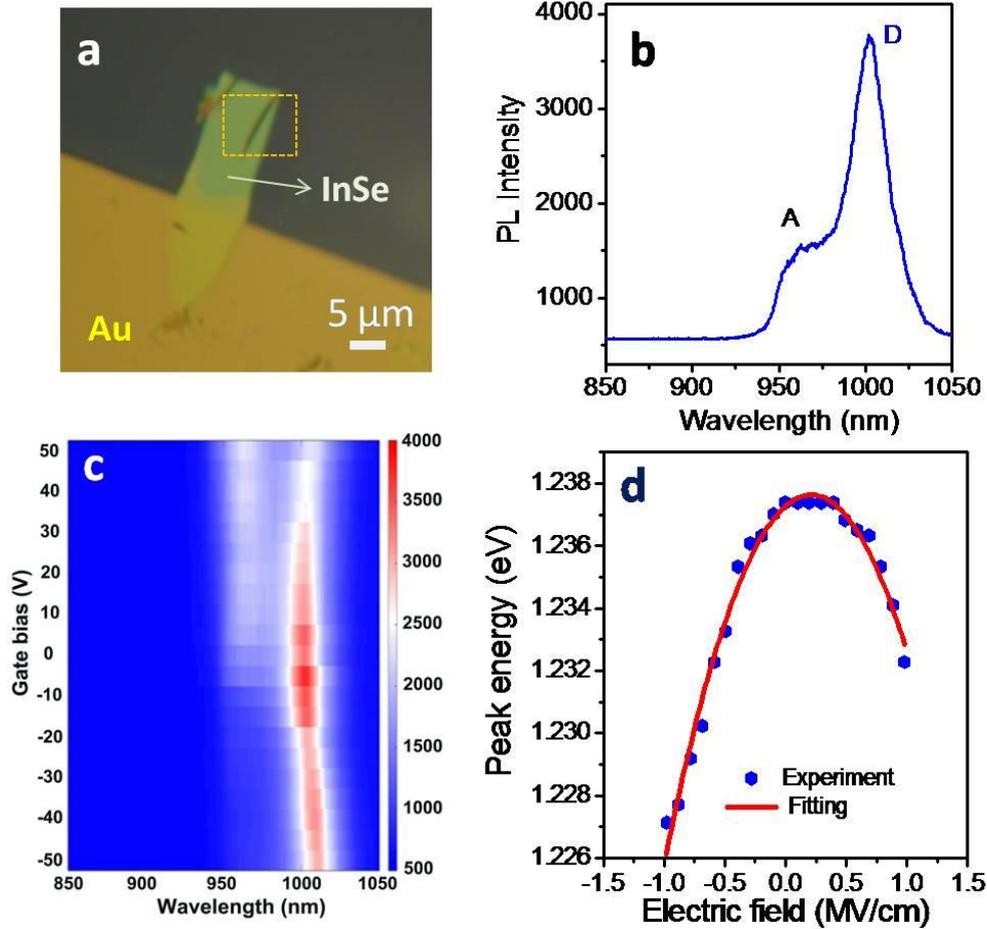

**Figure 4. Gate voltage tunable PL spectra.** (a) Optical image of the InSe device on which we performed the PL measurements at different gate voltages. (b) Typical PL spectra from the InSe ridge at 77 K. (c) PL intensity of the InSe ridge as a function of the emission wavelength and gate voltage at 77 K. (d) The new PL peak position (emission photon energy) from the InSe ridge as a function of the electric field at 77 K.

The localized nature of the defect would suggest a sensitive dependence on the electrical field through the Stark effect.[37-39] We thus fabricated InSe devices and used a back-gate voltage to apply an out-of-plane electrical field to investigate the PL change from the InSe ridge. For a typical device as shown in Fig. 4a, the PL spectrum from the ridge at 77 K with no gate voltage applied is shown in Fig. 4b, in which both the intrinsic exciton and defect emission are observed. The gate-voltage dependent PL spectra from the InSe ridge is shown in a color plot in Fig. 4c (also S7 in SI). As shown in Fig. 4c, in addtion to the enhanced PL amplitude, the new PL peak exhibits two features which are qualitatively disctinct from the intrinsic exciton peak: (1) the intensity of intrinsic exciton peak quenches from gate voltage –40 V to ~ 0 V, while the maximum intensity of the new PL peak is located at gate voltage ~ 0 V and remains strong from –50 to 30 V. (2) the intrinsic exciton peak position remains roughly at the same wavelength, while the new PL peak position sensitively depends on the gate voltage applied.

Both features can be explained with the interpretation that the new PL peak originates from localized defects. First, the quenching of the exciton PL has been widely observed in gated monolayer TMDCs devices, in which the addition of free carrier significantly enhances the non-radiative channels.[43,44] Localized defects, however, are less sensitive to the presence of free carriers. Second, it has been shown in previous reports that the Stark effect is enhanced in quantum confined systems.[37,38,40] Suggested by the previously extracted activation energies, the defects are much more localized than the exciton and hence the PL is more sensitive to the electrical field.

We can quantitatively understand the defect dipole moment by investigating the PL shift as a function of the gate voltage dependence (Fig. 4c). To achieve this, we convert the gate voltage to the electrical field using the 300 nm thickness of the oxide layer with a simple capacitor model. Due to the complicated geometry of ridge, the extracted electrical field will be an estimate of the upper bound. The detailed emission photon energy from defects as a function of the external electrical field is plotted in Fig. 4d. The dependence of the defect PL peak on the electric field can be fitted with a quadratic function: [39,40]

$$E(F) = E_0 - pF - \beta F^2, (2)$$

where $E_0$, $p$, $F$ and $\beta$ are the defect emission energy without electrical field, the permanent dipole moment, electric field and exciton polarizability. The linear and quadratic terms are the contributions from the permanent dipole moment and the electrical-field-induced dipole moment, respectively. The fitting shown in Fig. 4d agrees well with the experimental data with $p = -1.63$ D, and $\beta = 3.886 \times 10^{-8}$ $D$m/V (1 D = $3.33 \times 10^{-30}$ C·m). The permanent dipole moment $p$ of the defect states in InSe is comparable to that in a nitrogen-vacancy (NV) defect center (typically between −1.5 to 1.5 D),[40] and monolayer or few-layer $MoS_2$ (~ 1.35 D).[41,42] However, the polarizability of defect state in InSe is almost 6 time larger than that in monolayer $MoS_2$ ($0.58 \times 10^{-8}$ $D$m/V), indicating stronger confinement of the defect state than that in $MoS_2$. This strong confinement renders sensitive electrical field dependence of the PL emission, which can be explored for tunable optoelectronic devices. Additional data on Stark effect can be found in Fig. S19 in SI.

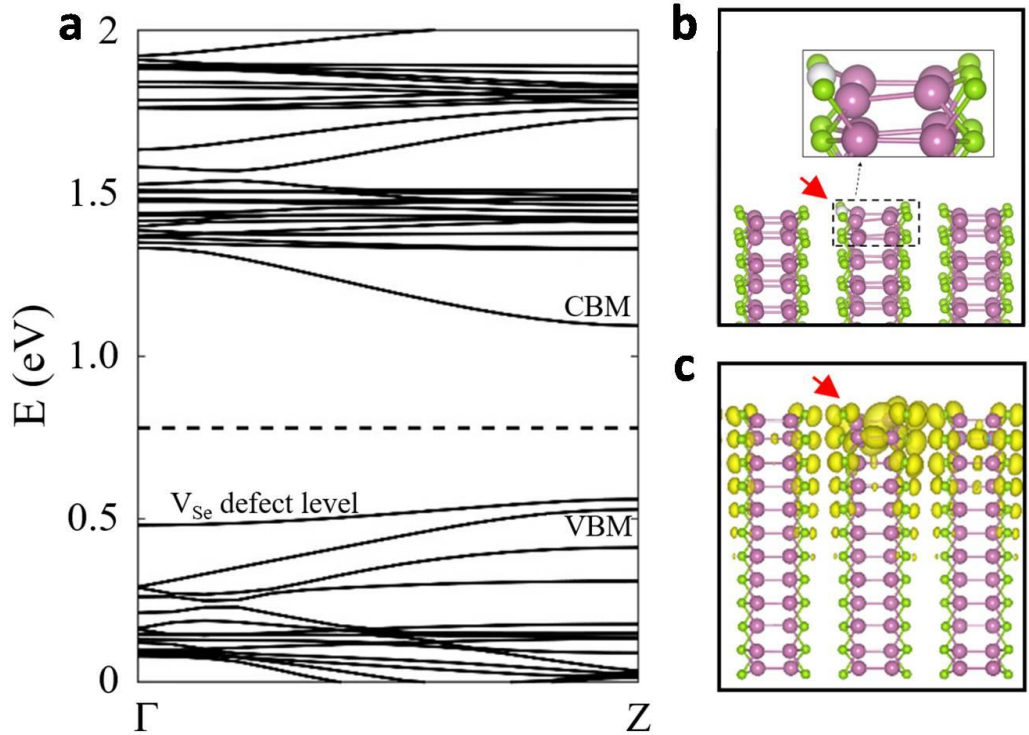

**Figure 5. Theoretical Calculation of the Se vacancy defect on the InSe armchair edge.** (a) Band structure of a defect supercell containing a single Se vacancy on the armchair terminated edge of InSe, and the Fermi energy is indicated by the dashed line and the band associated with the Se defect level is shown in red. (b) Atomic structure of the Se vacancy on the armchair terminated surface of InSe. Red and green spheres represent In and Se atoms, respectively, and the location of the Se vacancy is depicted with a white sphere and indicated by the arrow. (c) An isosurface of the charge density of the defect state associated with the vacancy, indicating the defect level is spatially localized at the InSe edge, in the vicinity of the vacancy.

According to the previous discussion, a fracture has to form in order to release the large strain of approximately 8.8%.[34,35] We further confirmed that the natural edge of the InSe flake does not give rise to the additional PL peaks, but the edge of a freshly scratched flake showed similar PL spectra as that from the InSe ridge (see SI). This suggest that the additional PL might originate from the defects formed due to the mechanical break. To determine a likely defect candidate, we have performed first-principles density functional theory calculations of point defects at the edge of InSe and find that the Se vacancy ($V_{Se}$) is a prime candidate. The formation energy of the $V_{Se}$ defect along the armchair edge, the structure of which is depicted in Fig. 5b, is found to be 0.02 eV under In-rich conditions, indicating that high concentrations are thermodynamically favorable at room temperature. Furthermore, examination of the bandstructure associated with this defect, shown in Fig. 5a, reveals the presence of a relatively flat occupied shallow acceptor state. An isosurface of the charge density associated with this defect state is depicted in Fig. 5c, where it can be seen that the defect state localizes to the vacancy at the edge region. As this defect level lies above the valence band maximum, it serves as a trap for excitons, wherein the hole associated with an exciton can lower its energy by approximately 30 meV by occupying the $V_{Se}$ defect state. This reduction in exciton energy is consistent with the observed red-shift in the PL associated with D-peak, which is observed in the vicinity of the ridge. Furthermore, the $V_{Se}$ trapped exciton model for the observed defect peak would also explain the observed linear Stark effect. As the hole localizes near the edge, this naturally causes anisotropy in the exciton wavefunction, wherein the electron, which is not as localized as the hole, extends further into the bulk region resulting in a sizeable permanent dipole moment. Hence, not only are Se vacancies likely to form at the fracture, but their presence would lead to an additional red shifted excitonic peak at low temperature which is sensitive to the gate voltage. While the quantitative agreement of the defect exciton binding is most likely fortuitous, the shallow nature of $V_{Se}$ at the edge, combined with its low formation energy, presence of linear Stark effect, and its likelihood to form under mechanical scratch, all point toward $V_{Se}$ at the edge as a likely candidate for the origin of the observed D-peak.

**Conclusions**

In summary, we have identified a unique geometric ridge state in 2D InSe flake. We have exploited the InSe ridge to bypass the selection rule and enhance optical absorption of normally incident light, and we have confirmed this enhancement through PL spectroscopy and scanning photocurrent microscopy. In addition, we observe greatly enhanced light emission from defect states on the ridge, almost two orders of magnitudes stronger than the intrinsic exciton PL. The emission from the defect states can also be greatly tunable by an external DC gate voltage through the Stark effect, thanks to the extremely localized nature of the defect state. The InSe ridge thus provides a new platform for defect engineering which can enable tunable optoelectronic devices in the near infrared regime.

**Experimental Methods**

Bulk InSe single crystals were synthesized using the vertical Bridgman method described before[36], which was used for mechanical exfoliation of InSe flakes with ridges. AFM topographies of the flakes were measured by atomic force microscopy (AFM, Bruker Dimension Icon-PT) operating in tapping mode for the angle and height determination. Mirco-PL measurements were performed at room temperature and low temperature using a home-built confocal micro-PL spectroscopy setup with the 532 nm laser excitation. PL mapping at room temperature was performed by Renishaw Invia micro-Raman system with the 532 nm laser excitation. Devices were fabricated by a shadow mask technique, followed by e-beam evaporation of 5 nm Ti/100 nm Au electrodes.

SPCM measurements were described previously.[36] A laser beam (a CW or supercontinuum laser) was focused on the samples with a diffraction-limited spot and scanned on the devices mounted on a piezo-stage, while the photocurrent at different positions was recorded by a Data Acquisition Card (DAC) coupled with a current amplifier. For the excitation wavelength dependent photocurrent measurements, a supercontinuum laser (Fianium) with the wavelength ranging from 700 to 1025 nm was used to illuminate the ridge and flat region on the devices.

Our first-principles calculations were based on density functional theory with the Perdew-Burke-Ernzerhof (PBE) approximation to the exchange-correlation function. The interactions between ion cores and valence electrons were described by the projector augmented wave (PAW) potentials as implemented in the VASP code. Plane waves with a kinetic energy cutoff of 240 eV were used as the basis set. The calculations were carried out in periodic supercells. The convergence criterion for all atoms is 0.025 eV/Å. A $1 \times 1 \times 2$ k-point set was used for the BZ integration.

**Acknowledgment**

S.-F. Shi acknowledges support from the AFOSR under Grant FA9550-18-1-0312. HW and SBZ were supported by the U.S. DOE Grant No. DESC0002623. DW acknowledges support from the NSF under Award No. EFMA-1542798. The supercomputer time was sponsored by NERSC under DOE Contract No. DE-AC02-05CH11231, and the CCI at RPI are also acknowledged. The device fabrication was supported by Micro and Nanofabrication Clean Room (MNCR), operated by the Center for Materials, Devices, and Integrated Systems (cMDIS) at Rensselaer Polytechnic Institute (RPI). Cheng-Yan Xu acknowledges the support from National Natural Science Foundation of China (No. 51572057).

**Supporting Information**

The supporting Information is available free of charge on the ACS Publications website.

It includes angle distribution of ridges, PL mapping of InSe flakes with ridges at room temperature, photocurrent mapping of InSe flakes with ridges at 77 K, defect emission of InSe flakes with ridges at 77 K and its origin discussion, discussion of the out-of-plane dipole in 2D InSe and the Arrhenius equation.

**Author Information**

§ Y.L, T.W and H.W contributed equally.

* Corresponding Authors:
Email: shis2@rpi.edu, zhangs9@rpi.edu, cy_xu@hit.edu.cn

ORCID:
Han Wang: 0000-0002-7476-6012
Christian Wetzel: 0000-0002-6055-0990
Cheng-Yan Xu: 0000-0002-7835-6635

ShengBai Zhang: 0000-0003-0833-5860
Damien West: 0000-0002-4970-3968
Su-Fei Shi: 0000-0001-5158-805X

**Notes**

The authors declare no competing financial interest.

TOC

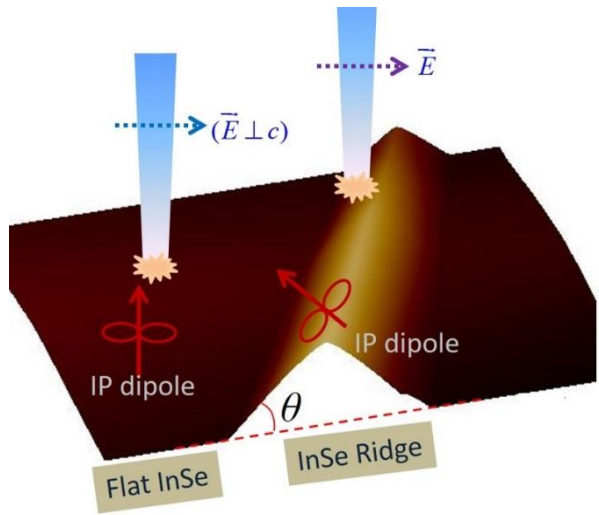 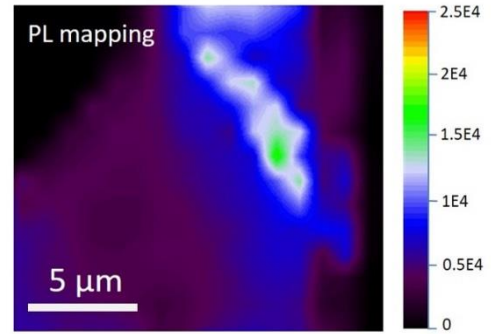